\documentclass[letter]{aa} 
\usepackage{graphicx}
\usepackage{txfonts}
\usepackage{natbib}

\newcommand{\tablenotea}[1]{\parbox{8.6cm}{ \indent
\footnotesize{\textsc{Note.--}~#1}}}

\newcommand{\jms}{J.~Mol.~Spectr.}                
\newcommand{\jphysb}{J.~Phys.~B:~Mol.~Opt.~Phys.} 
\newcommand{\jmst}{J.~Mol.~Struct.}               
\newcommand{\pccp}{Phys.~Chem.~Chem.~Phys.}       

\begin{document}
\title{Tentative detection of phosphine in IRC +10216 \thanks{Based on
observations carried out with the IRAM 30-meter telescope and the
Caltech Submillimeter Observatory (CSO). IRAM is supported by
INSU/CNRS (France), MPG (Germany) and IGN (Spain). The CSO is
operated by the California Institute of Technology under funding
from the National Science Foundation, Grant No. AST-0540882.}}
\titlerunning{Tentative detection of PH$_3$ in IRC +10216}
\authorrunning{Ag\'undez et al.}

\author{M. Ag\'undez\inst{1}, J. Cernicharo\inst{1}, J. R. Pardo\inst{1}, M. Gu\'elin\inst{2},
T. G. Phillips\inst{3}}

\offprints{M. Ag\'undez}

\institute{Departamento de Astrof\'isica Molecular e Infrarroja,
Instituto de Estructura de la Materia, CSIC, Serrano 121, 28006
Madrid, Spain; \email{marce@damir.iem.csic.es,
cerni@damir.iem.csic.es,pardo@damir.iem.csic.es} \and Institut de
Radioastronomie Millim\'etrique, 300 rue de la Piscine, 38406 St.
Martin d'H\'eres, and LERMA/\'Ecole Normale Sup\'erieure, 24 rue
Lhomond, 75231 Paris, France; \email{guelin@iram.fr} \and
California Institute of Technology, Downs Laboratory of Physics
320-47, Pasadena, CA 91125, USA; \email{tgp@submm.caltech.edu}}

\date{Received ; accepted }


\abstract
{}
{The J$_{\rm K}$ = 1$_0$-0$_0$ rotational transition of phosphine
(PH$_3$) at 267 GHz has been tentatively identified with a T$_{\rm
MB}$ $\sim$40 mK spectral line observed with the IRAM 30-m
telescope in the C-star envelope IRC +10216.}
{A radiative transfer model has been used to fit the observed line
profile.}
{The derived PH$_3$ abundance relative to H$_2$ is 6 $\times$
10$^{-9}$, although it may have a large uncertainty due to the
lack of knowledge about the spatial distribution of this species.
If our identification is correct, it implies that PH$_3$ has a
similar abundance to that reported for HCP in this source, and
that these two molecules (HCP and PH$_3$) together take up about 5
\% of phosphorus in IRC +10216. The abundance of PH$_3$, as that
of other hydrides in this source, is not well explained by
conventional gas phase LTE and non-LTE chemical models, and may
imply formation on grain surfaces.}
{}


\maketitle
%

\section{Introduction}

Of the nearly 150 molecules detected so far in interstellar and
circumstellar media, around 3/4 can be formed from just four
elements (H, C, N, and O). The remaining 1/4 contain metals (Na,
K, Al, Mg, and Fe), halogens (F and Cl), and to a large extent the
second-row elements Si, P, and S. The scarcity of molecules
containing the second-row elements Si, P, and S, compared to their
first-row analogues C, N, and O, reflects on one hand a lower
cosmic abundance (Si/C $\sim$ 1/8, P/N $\sim$ 1/300, and S/O
$\sim$ 1/30; \citealt{asp05}) and on the other hand important
chemical differences. Second-row elements have generally a larger
refractory character, i.e. they tend to form solid condensates and
deplete from the gas phase \citep{fie74,tur91,lod99}. Also,
chemical bonds formed by first-row elements are generally stronger
and the resulting molecules are more stable than the corresponding
second-row analogues. For example, the high stability of C-C bonds
allows to form a large variety of organic molecules, while a
Si-based chemistry is much more limited.

Concerning phosphorus, the very limited number of interstellar and
circumstellar P-bearing molecules known for a long time, just PN
and CP \citep{tur87a,ziu87,gue90} in contrast with the nearly 50
N-bearing molecules found, may have been related to a sort of
mixing of all the factors mentioned above. First, the more than a
hundred fold decrease in the cosmic abundance of P compared to
that of N. Second, the likely strong depletion of phosphorus
suggested by observations of some P-bearing species \citep{tur90}.
And lastly, some important differences between the chemistry of P
and that of N, that are indicated by, for example, the non
reactivity of PH$_n^+$ ($n$ = 0,3) ions with H$_2$ as opposed to
the case of NH$_n^+$ \citep{tho84,ada90}.

In recent years, however, we have seen the discoveries, mostly in
circumstellar media, of new phosphorus compounds such as HCP, PO
and CCP \citep{agu07,mil08,ten07,hal08}, which permit to discuss
the chemistry of phosphorus on a wider observational basis. In
this letter we present the tentative detection of PH$_3$, the
phosphorus analogue of NH$_3$, in the carbon-rich circumstellar
envelope IRC +10216. This species, known to be abundant in the
atmospheres of the giant gaseous planets Jupiter and Saturn
\citep{wei96}, has never been observed outside the Solar System.

\begin{figure}
\includegraphics[angle=0,scale=.66]{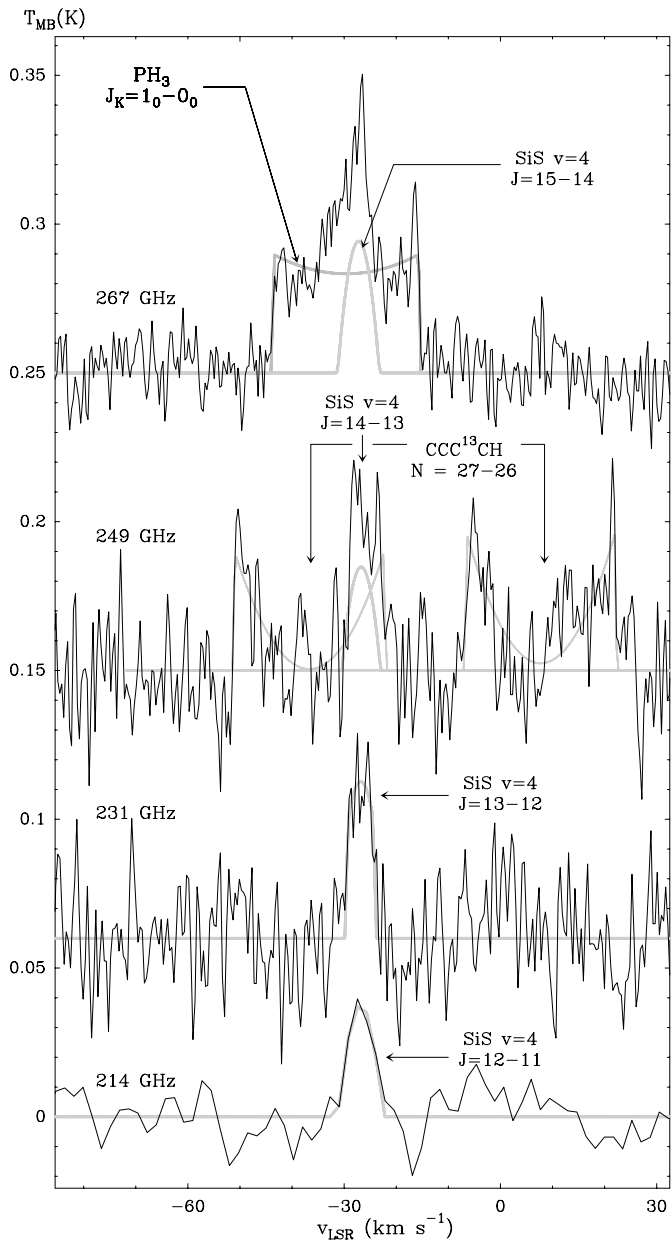}
\caption{Spectra of IRC +10216 observed with the IRAM 30-m
telescope showing on top the J$_{\rm K}$ = 1$_0$-0$_0$ line of
PH$_3$ blended with a narrower line assigned to the J =15-14 line
of SiS v = 4. Shown below are the immediate lower rotational
transitions of SiS v = 4. Fits to the line profiles using the
CLASS \emph{shell} method are shown as grey thick lines.}
\label{fig-30m-spectra} \vspace{-0.1cm}
\end{figure}

\section{Observations and results}

The phosphine molecule, PH$_3$, is an oblate symmetric top, thus
its rotational levels are given by two quantum numbers (J, K), and
radiative transitions are only allowed within levels of the same K
ladder ($\Delta$J = 1, $\Delta$K = 0). The K ladders are grouped
into two distinct forms: ortho (K = 3n, n an integer) and para (K
$\neq$ 3n) between which both radiative and collisional
transitions are severely forbidden. Its rotational spectrum has
been extensively investigated in the laboratory, allowing for the
very weak "forbidden" transitions ($\Delta$J = 0, $\Delta$K =
$\pm$3) to be measured and for the hyperfine structure due to the
$^1$H and the $^{31}$P nuclear spins to be resolved (see
\citealt{caz06} and references therein). In contrast to NH$_3$, no
evidence of inversion doubling has been found in the case of
PH$_3$. The electric dipole moment has been measured as 0.57395
$\pm$0.0003 D \citep{dav71}.

Following the recent detection of HCP in IRC +10216 \citep{agu07},
it was speculated that PH$_3$ might be detectable in this source
if the PH$_3$/HCP abundance ratio is similar to the NH$_3$/HCN
one. Prompted by this hypothesis we searched for the fundamental
ortho-PH$_3$ 1$_0$-0$_0$ line, at 267 GHz, with the IRAM 30-m
telescope. Preliminary observations with a low spectral
resolution, 1.25 MHz, were done in 2007 May and we found
significant emission at the frequency of the PH$_3$ 1$_0$-0$_0$
line and with the expected linewidth (v$_{\rm exp}$ = 14.5 km
s$^{-1}$ for most of the molecular lines in IRC +10216;
\citealt{cer00}). The line, however, showed a profile unusual for
IRC +10216: neither really U-shaped nor flat-topped. Encouraged by
this result we returned to the 30-m telescope in 2008 February and
April to re-observe this transition with a 4 times higher spectral
resolution (320 kHz). Two SIS receivers operating at 1 mm were
used simultaneously with upper side band rejections of $\sim$ 10
dB. The local oscillator was shifted by 80 MHz to identify any
contribution from the image sideband. An autocorrelator was used
as backend to provide the required spectral resolution of 320 kHz.
The secondary mirror was wobbled by $\pm$90$''$ at a rate of 0.5
Hz. The pointing and focus of the telescope were checked every 1-2
hours on Saturn, which was closer than 10$^{\circ}$ to IRC +10216.
Here we express the intensity scale in units of main beam
brightness temperature T$_{\rm MB}$. The parameter B$_{\rm
eff}$/F$_{\rm eff}$, used to convert T$_{\rm A}^*$ into T$_{\rm
MB}$, is 0.51 and the beam size is 9'' at 267 GHz for the 30-m.

\begin{table} \caption{Observed
line parameters in IRC +10216} \label{table-lineparameters}
\centering
\begin{tabular}{lcccc}
\hline \hline
\multicolumn{1}{c}{}           & \multicolumn{1}{c}{Cal. Freq.} & \multicolumn{1}{c}{Obs. Freq.} & \multicolumn{1}{c}{v$_{\rm exp}$$^a$}     & \multicolumn{1}{c}{$\int$$T_{\rm MB}$dv} \\
\multicolumn{1}{c}{Transition} & \multicolumn{1}{c}{(MHz)}      & \multicolumn{1}{c}{(MHz)}      & \multicolumn{1}{c}{(km/s)} & \multicolumn{1}{c}{(K km/s)} \\
\hline
\multicolumn{5}{l}{PH$_3$} \\
J$_{\rm K}$ = 1$_0$-0$_0$  & 266944.514  &  266944.5(3)    & 14.1(3)  & 1.002(40) \\
\hline
\multicolumn{5}{l}{SiS v = 4} \\
J = 15-14          & 266941.754  &  266942.4(7)    & 4.0(7)   & 0.237(40) \\
J = 14-13          & 249155.372  &  249155.6(4)    & 3.5(4)   & 0.172(40) \\
J = 13-12          & 231366.976  &  231367.2(3)    & 2.7(3)   & 0.256(30) \\
J = 12-11          & 213576.710  &  213576.9(5)    & 3.5(5)   & 0.207(40) \\
\hline
\end{tabular}
\tablenotea{Number in parentheses are 1$\sigma$ uncertainties in
units of the last digits. Observed frequencies are given in the
rest frame of IRC +10216 assuming a systemic velocity of v$_{\rm
LSR}$ = -26.5 km s$^{-1}$.\\
$^a$ v$_{\rm exp}$ is the half width
at zero level.}
\end{table}

The resulting spectrum at 267 GHz is shown on top of
Fig.~\ref{fig-30m-spectra}. The line profile indicates the blend
of a normal cusped line of width $\sim$29 km s$^{-1}$,
characteristic of optically thin lines arising in the outer
envelope (e.g. C$_2$H), with a narrower ($\sim$6 km s$^{-1}$)
sharply peaked line, characteristic of vibrationally excited lines
arising close to the star. The latter line was soon identified as
the J = 15-14 transition of SiS in its v = 4 vibrational state,
based on SiS laboratory spectroscopic data \citep{san03,mul07}.
Emission of SiS in vibrationally excited states up to v = 3 has
been reported toward IRC +10216 \citep{tur87b,fon06,cer00}. The
linewidths of vibrationally excited SiS are unusually narrow
implying that the emission arises from the innermost envelope
where the gas has not yet reached the terminal expansion velocity
of 14.5 km s$^{-1}$. To constrain the possible contribution of the
J = 15-14 line of SiS v = 4 to the 267 GHz line, we observed the
immediate previous J transitions. In Fig.~\ref{fig-30m-spectra} we
show the J = 14-13 and J = 13-12 lines of SiS in the v = 4 state
(both observed with a spectral resolution of 320 kHz at the time
of the 267 GHz observations) and the J = 12-11 line (observed in a
previous run in 2005 January with a spectral resolution of 1.25
MHz).

Except for the J= 14-13 line of SiS v = 4, which is partially
blended with a fine structure component of the N = 27-26
transition of CCC$^{13}$CH, the observations clearly show that SiS
v = 4 lines are narrow, with expansion velocities of about 3 km
s$^{-1}$ (see Table~\ref{table-lineparameters}). With this in
mind, the observed emission feature at 267 GHz has been fitted by
two line components, a narrow one corresponding to the J = 15-14
transition of SiS v = 4 and a wider line whose width, v$_{\rm
exp}$ = 14.1 $\pm$0.3 km s$^{-1}$, agrees with the expansion
velocity of 14.5 km s$^{-1}$ in IRC +10216, and whose center rest
frequency, 266944.5 $\pm$0.3 MHz, is in very good agreement with
the laboratory frequency of the 1$_0$-0$_0$ line of PH$_3$ (see
Table~\ref{table-lineparameters}). The hyperfine structure is not
resolved as the components are separated by less than 0.2 km
s$^{-1}$ in velocity \citep{caz06}, which is smaller than the
spectral resolution. The good agreement between the observed and
laboratory frequencies is the strongest evidence of PH$_3$
detection in IRC +10216.

Besides the SiS v = 4 line, there are some other lines with
frequencies close to that of PH$_3$ 1$_0$-0$_0$. Most of them,
such as SO$_2$ 30$_{9,21}$-31$_{8,24}$ at 266943.344 MHz or
CH$_3$CH$_2$CN 15$_{4,12}$-15$_{2,13}$ at 266951.639 MHz, are
ruled out as likely contributors since many other lines of these
species should have been detected. A more plausible species is
HC$_3$N in the $\nu_7$ = 4 vibrational state, whose transition J =
29-28 $\ell$ = 0e lies at 266943.313 MHz \citep{mbo00}. Although
several lines of HC$_3$N in the vibrational excited state $\nu_7$
= 1 have been observed in IRC +10216 \citep{cer00}, we rule out
that HC$_3$N $\nu_7$ = 4 is the main contributor to the 267 GHz
emission based on the upper limit of T$_{\rm MB}$ $<$ 0.01 K that
we have from our 30-m data archive for the lower-J transition J =
26-25 $\ell$ = 0e transition at 239370.171 MHz.

We tried to confirm the identification of PH$_3$ by observing
other transitions. The J = 2-1 line at 534 GHz is not reachable
from the ground due to severe atmospheric absorption. We, thus,
searched for the J = 3-2 transition at 800 GHz with the Caltech
Submillimeter Observatory (CSO). The observations were carried out
in 2008 January using the chopping secondary mode with a throw of
$\pm$90'' at a rate of 1.2 Hz. The SIS receiver was tuned in
double sideband and a Fast Fourier Transform Spectrometer was used
as backend to provide a spectral resolution of 0.12 MHz. The
pointing of the telescope was checked on Saturn. The beam size of
the CSO at 800 GHz is 11.5'' and the beam efficiency is 0.28. In
spite of the good atmospheric conditions (zenith sky opacity at
225 GHz was 0.04-0.08), the high opacity of the atmosphere at 800
GHz (T$_{\rm sys}$ ranged from 3000 K to 8000 K) did not allowed
to reach a noise level low enough to confirm or discard the
presence of PH$_3$ (see top panel in Fig.~\ref{fig-lvg-lines}).

\section{Analysis and discussion}

To interpret the observations we have computed line profiles by
means of excitation and radiative transfer calculations based in
the Large Velocity Gradient (LVG) formalism \citep{cas70}, coupled
to the spectral catalog of J. Cernicharo \citep{cer00}. We
consider separately both the ortho (o-PH$_3$) and para (p-PH$_3$)
species of phosphine (the ortho-to-para ratio is assumed to be 1,
the statistical value). The energy levels and transition
frequencies are computed from the rotational constants reported by
\citet{caz06}. We include rotational levels in the ground
vibrational state up to J$_{\rm K}$ = 7$_6$ for o-PH$_3$ and up to
J$_K$ = 5$_{\rm 5}$ for p-PH$_3$. As rate coefficients for
collisional de-excitation of o-PH$_3$(p-PH$_3$) with H$_2$ and He,
we have adopted those computed for collisions of
o-NH$_3$(p-NH$_3$) with p-H$_2$ \citep{dan88} and He \citep{mac05}
respectively, properly corrected to the case in which inversion
doubling is not resolved.

\begin{figure}
\includegraphics[angle=0,scale=.63]{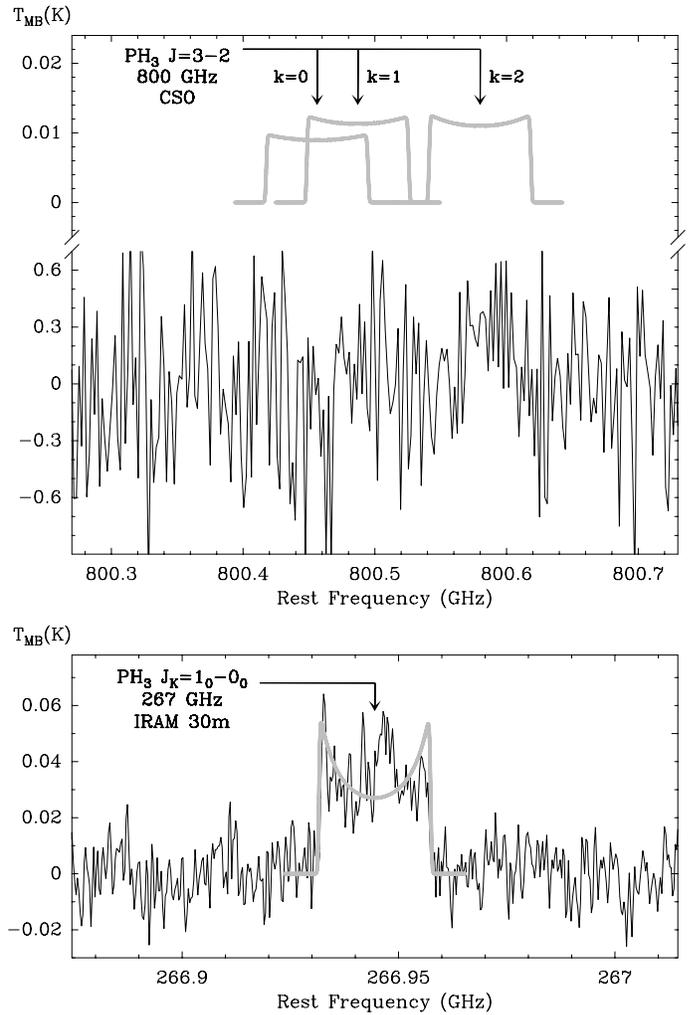}
\caption{The lower panel shows the PH$_3$ J$_{\rm K}$ =
1$_0$-0$_0$ line at 267 GHz as observed with the IRAM 30-m
telescope, in which the fit to the SiS v = 4 J = 15-14 line has
been subtracted. The upper panel shows the observation at 800 GHz
with the CSO smoothed to a spectral resolution of 2 MHz. The grey
thick lines in both panels correspond to the line profiles given
by the LVG model.} \label{fig-lvg-lines} \vspace{-0.1cm}
\end{figure}

We assume a distance to IRC +10216 of 150 pc and simulate the
circumstellar envelope as a spherically distributed gas expanding
at a constant velocity of 14.5 km s$^{-1}$. The gas density and
temperature radial profiles are taken from \citet{agu06}. PH$_3$
is assumed to be distributed between an inner radius R$_{\rm in}$
and an outer radius R$_{\rm out}$, with a constant abundance,
$x$(PH$_3$), relative to H$_2$. We take R$_{\rm in}$ = 1.3
$\times$ 10$^{15}$ cm (about 20 stellar radii), the value adopted
by \citet{has06} for NH$_3$. The values of R$_{\rm out}$ and
$x$(PH$_3$) are varied until obtaining the best fit to the
observed PH$_3$ 1$_0$-0$_0$ line profile. Since both gas density
and temperature vary greatly with radius, we have divided the
envelope into various shells and solved for the level populations
in each shell independently of the others. We then compute the
emergent intensity and weight it with the main beam of the
selected telescope.

In the bottom panel of Fig.~\ref{fig-lvg-lines} we plot the
observed 1$_0$-0$_0$ PH$_3$ line profile, obtained by subtracting
the fit to the SiS v = 4 blended line. Also shown is the profile
resulting from our best LVG model, whose parameters are R$_{\rm
out}$ = 2 $\times$ 10$^{16}$ cm and $x$(PH$_3$) = 6 $\times$
10$^{-9}$. The model indicates that the upper level involved in
the 1$_0$-0$_0$ PH$_3$ line is mostly populated by collisions in
the inner 4 $\times$ 10$^{15}$ cm, which corresponds to an angular
diameter of about 4'' and is thus spatially diluted in the 9''
beam of the 30-m telescope. The top panel of
Fig.~\ref{fig-lvg-lines} shows the corresponding prediction for
the PH$_3$ J = 3-2 line profiles as observed with the CSO. It is
seen that the expected intensity is well below the noise of the
observed spectrum and thus is consistent with the IRAM 30-m result
for the J = 1-0 line. The spatial dilution is also important for
the CSO observations and plays against our efforts to detect the J
= 3-2 lines.

It should be noted that the abundance and distribution of PH$_3$
derived are very uncertain. Since we have just one single PH$_3$
line observed, there exists a considerable degeneracy between
models with different values of R$_{\rm in}$, R$_{\rm out}$ and
$x$(PH$_3$). Moreover, infrared pumping to excited vibrational
states, not considered in our model, may play an important role in
the excitation of the rotational levels in the ground vibrational
state. PH$_3$ has indeed many vibrational bands in the spectral
region around 10 $\mu$m, a wavelength at which the central source
has its maximum flux \citep{cer99}.

Confirmation of our tentative detection may rely on further
observations of other PH$_3$ transitions. The J = 2-1 and J= 3-2
lines, at 534 GHz and 800 GHz respectively are observable with the
Herschel Space Observatory (HSO), although our LVG model predicts
intensities somewhat weaker than with the CSO due to the larger
dilution effect. In the case of the Atacama Large Millimeter Array
(ALMA) the high angular resolution that it will provide, better
than 3'' at 800 GHz, will perfectly fit with the expected size of
the J = 3-2 emission. The predictions indicate main beam
brightness temperatures of about 1 K, which provide a good
opportunity of detection in spite of the high atmospheric opacity
at this frequency. We note that these predictions are based on our
best LVG model, but the J = 2-1 and J = 3-2 lines could be more
intense than expected if infrared pumping is playing an important
role, or if PH$_3$ is present in hotter regions, inner than 1.3
$\times$ 10$^{15}$ cm.

The derived abundance of PH$_3$, $x$(PH$_3$) = 6 $\times$
10$^{-9}$, implies that it takes up about 1.3 \% of the available
phosphorus. The PH$_3$/HCP ratio is then found to be 1/2.3
\citep{agu07}, noticeably higher than the NH$_3$/HCN ratio, that
is about 1/50 in IRC +10216 \citep{has06,fon07}. The formation of
PH$_3$ in the gas phase is difficult to explain. Its LTE abundance
in the inner envelope is rather low, $<$ 10$^{-12}$, and gas phase
reactions yield no net formation in the outer envelope
\citep{agu07}. Besides PH$_3$, other hydrides such as NH$_3$,
CH$_4$, H$_2$O, SiH$_4$, and H$_2$S are observed in IRC +10216
with relatively high abundances \citep{kea93,has06,agu06}. Many of
them are usually assumed as parent molecules, i.e. formed in the
inner envelope, in most chemical models of IRC +10216 (e.g.
\citealt{mac01,agu06}). However, chemical equilibrium
calculations, similar to those reported in \citet{agu07}, indicate
abundances much lower than observed, except perhaps for CH$_4$. A
widely invoked explanation, when gas phase chemistry fails to
explain an observed abundance, is that of grain surface reactions.
In the case of hydrides such as PH$_3$, a likely formation process
is the direct hydrogenation of the heavy atom taking place on
grain surfaces.

\section{Conclusions}

We have tentatively detected PH$_3$ in IRC +10216 through its
J$_{\rm K}$ = 1$_0$-0$_0$ transition at 267 GHz. The derived
abundance relative to H$_2$ is 6 $\times$ 10$^{-9}$. Although with
a considerable uncertainty, this value is similar to the HCP
abundance found in this source \citep{agu07}. These two species,
HCP and PH$_3$, would then take up about 5 \% of the phosphorus in
IRC +10216. The formation of PH$_3$, unlike that of HCP, is
difficult to explain in the gas phase and could occur on grain
surfaces. It remains a target for the future to confirm this
tentative detection by observing the J = 3-2 transition, at 800
GHz, with the ALMA facility. Also, further observations at 267 GHz
in other sources such as CRL 2688, where HCP has been also
detected \citep{mil08}, will be of great interest to support this
tentative detection and to understand the chemistry behind it.

\begin{acknowledgements}
We thank the IRAM staff, especially C. Thum, for their kindness
during the 30-m observations. We also acknowledge funding support
from Spanish MEC trough grants AYA2006-14876 and ESP2004-665, and
from Spanish CAM under PRICIT project S-0505/ESP-0237 (ASTROCAM).
MA also acknowledges grant AP2003-4619 from Spanish MEC.
\end{acknowledgements}

\end{document}